\documentstyle[prl,aps,twocolumn]{revtex}

\begin{document}

\title{Effective Medium Theory of the Optical Properties of Aligned Carbon
Nanotubes}

\author{F.J. Garc\'{\i}a-Vidal \cite{ic}, J.M. Pitarke \cite{bi} 
and J.B. Pendry}

\address{
Condensed Matter Theory Group, \\
The Blackett Laboratory, Imperial College, \\
London SW7 2BZ (UK).}

\maketitle

\begin{abstract}
We present an effective medium theory in order to analyze 
the reported optical properties of aligned carbon nanotube films 
(W.A. de Heer {\it et al.}, Science {\bf 268}, 845 (1995)). 
This methodology is based on photonic band structure 
calculations and allows treatment of complex media consisting of 
particles that interact strongly. We also develop a simple 
Maxwell-Garnett type approach for studying this system. 
In comparing the results of both mean field theories, we demonstrate that 
the inclusion of the full electromagnetic coupling between the 
nanotubes, as our numerical scheme does, is necessary for 
a complete explanation of the experimental data. 
\newline
\end{abstract}

PACS numbers: 78.66.Sq, 41.20.Jb, 61.46.+w, 73.20.Mf  \newline

Since the discovery of tubular fullerenes in 1991 \cite{Ijima}, several
experimental and theoretical studies of the unique properties
of these materials have been carried out. Despite the short time
since the first carbon nanotube was synthesised, possible
applications of these new materials have already been reported. 
Nanotubes
can be used as atomic-scale field emitters \cite{waheer} or as
a pinning material in high-$T_c$ superconductors \cite{Foss} whereas 
it has been proposed that hollow tubes might serve as nanoscale 
molds.
The unique properties of these carbon tubes are derived from their
special topology and their nanometric dimensions.
Very recent measures of the conductivity of individual nanotubes 
\cite{langer} suggest that the geometry of these tubes plays 
a fundamental role in determining if the nanotube is 
metallic or semiconducting. From the mechanical point of 
view, nanotubes seem to be very strong and flexible at the 
same time due to the ability of sp$^2$ carbon atoms 
to rehybridize when the bonds are deformed.

These carbon
nanotubes also present very interesting optical properties.
Thin films of aligned carbon tubes are birefringent, reflecting
differences in the dielectric function for light polarized 
 along (s-polarization) and normal (p-polarization) to the
tubes \cite{waheer2}. Up to our knowledge, a complete theoretical 
analysis of the optical properties of these arrays of carbon 
nanotubes is still missing. 

In analyzing the propagation of 
electromagnetic waves in heterogeneous media,   
various effective medium 
approaches like the Maxwell-Garnett (MG) approximation \cite{Max} 
have been used to determine the effective dielectric constant 
of different composite structures. Very recently, 
the validity of the MG approach has been checked by comparing 
their results for periodic dielectric systems with the ones 
obtained in the long-wavelength limit of photonic band structure 
calculations \cite{Datta}.  For these structures, the MG results  
seem to agree reasonably well with their vector-wave counterparts for small 
filling ratios, but in the close-packing regime the inclusion of the 
full electromagnetic (EM) interaction between the different objects 
is unavoidable. Only numerical methods give the capacity to 
treat this more complex limit.   
 
Our idea is to apply a numerical methodology
originally constructed to study photonic materials and the propagation
of EM waves in complex media \cite{John} to analyze the
optical properties of these carbon nanotube films where the 
tubes form a very close-packed structure. Within this   
framework it is possible to develop an effective medium theory
in which the EM interaction between the constituents of the
complex system is fully included \cite{Luis}. 
We also want to develop a MG-like approach 
able to study arrays of 
carbon nanotubes and compare these results   
with our {\it exact} results, studying the 
range of validity of MG approximation for carbon based materials. 

Our model for a carbon nanotube film is shown in figure 1a. An
array of tubes, infinitely long in the $z$-direction, is
arranged on a square lattice. The diameter of the cylinders is
chosen to be $10$ nm, in accordance with experimental evidence 
\cite{waheer2}.
The distance between the nanotubes, $d$, will be varied throughout our
calculations, studying the dependence of the
optical properties of this structure on the separation
between the tubes.

Let us discuss first the dielectric model we are using for
describing the dielectric characteristics of an isolated
carbon nanotube. Carbon nanotubes are made of a number of
cylindrical shells of planar graphite. Graphite is highly
anisotropic and its optical response depends on the direction
of the electric field with respect to the axis normal to the
graphite sheets. Therefore, for a given frequency $\omega$, 
 it is necessary to distinguish
two different components, $\epsilon_{\parallel}(
\omega)$ and $\epsilon_{\perp}(\omega)$, in its dielectric 
function 
for the directions parallel and normal to the axis, respectively.
These components have been measured by several methods, leading to 
slightly different sets of data. For our calculations 
we have used the dielectric functions of graphite as 
tabulated in Ref. \cite{Palik}.
In transfering these dielectric constants to cylindrical
multishells we will assume that this material is
a dielectric continuum and locally identical 
to graphite. A similar transfer procedure has 
been previously applied to compute the UV absorption 
of multishell fullerenes \cite{Lucas}. 
At every point inside the nanotube we can write down
a local dielectric tensor that in cylindrical coordinates
(see fig 1b) takes the diagonal form:

\begin{equation}
\hat{\varepsilon} (\omega)=\varepsilon_{\perp}(\omega)
({\bf \theta}{\bf \theta}+{\bf z}{\bf z})+
\varepsilon_{\parallel}(\omega){\bf r}{\bf r}.
\end{equation}

In Cartesian coordinates, the tensor like character of the
dielectric function is more pronounced:

\begin{equation}
\hat{\varepsilon} (x,y,\omega)=\left( \begin{array}{ccc}
\frac{x^2}{r^2}\varepsilon_{\parallel}+
\frac{y^2}{r^2}\varepsilon_{\perp} &
\frac{xy}{r^2}(\varepsilon_{\parallel}-\varepsilon_{\perp}) &  0 \\
\frac{xy}{r^2}(\varepsilon_{\parallel}-\varepsilon_{\perp}) &
\frac{y^2}{r^2}\varepsilon_{\parallel}+
\frac{x^2}{r^2}\varepsilon_{\perp} & 0 \\
0 & 0 & \varepsilon_{\perp} \end{array} \right).
\end{equation}

Arc-generated multilayered nanotubes are generally hollow. 
Their inner radius distribution usually varies from 
$0.25$ to $2$ nm, being tubes with inner radius of 
$0.75$ to $1$ nm the most abundant ones in the samples \cite{ugarte}. 
The fundamental importance of the hollow 
character of {\it isolated} nanotubes in their dielectric 
response has been previously stressed \cite{Lucas2}.
However we have found that when nanotubes form close-packed 
structures, as in nanotube films, the presence of 
a hollow core plays a much less important role.
Moreover, our results indicate that in the strong coupling limit and 
for p-polarized light, the hollow core 
can be safely ignored if the ratio between the inner and the 
outer radii of the tubes is less than $0.4$. As the nanotubes 
present in the experiment \cite{waheer2} seem to fulfill this 
condition, for p-polarized light we will only show 
results for plain cylinders (inner radius equal to zero). 
A detailed discussion of the effect of 
a hollow core on the dielectric properties  
of aligned carbon nanotubes will be presented elsewhere \cite{txema}. 

If the nanostructured film could be replaced by an effective
homogeneous medium,  
a single pair of Bloch waves for 
each polarization, ($k_s(\omega)$,$-k_s(\omega)$) 
and ($k_p(\omega)$,-$k_p(\omega)$), should 
dominate the photonic band structure of the system. It is indeed 
the case that over the entire range of frequencies $\omega$, 
four EM waves are dominant and we can use  
the dispersion relation
of these Bloch waves to define the effective dielectric
constants for s and p polarizations 

\begin{equation}
\varepsilon_{eff}^{s}(\omega)=
\frac{c^2 k_{s}^{2}(\omega)}{\omega^2},\;\;\;\;\;
\varepsilon_{eff}^{p}(\omega)=
\frac{c^2 k_{p}^{2}(\omega)}{\omega^2},
\end{equation}
\noindent where $c$ is the speed of light.

In order to calculate the photonic band structure of the array of
carbon nanotubes shown in figure 1a,
we need to find solutions of the Maxwell equations
with a tensor dielectric function,

\begin{eqnarray}
\nabla \times  {\bf E}&=&-\mu _0 \mu \;\partial 
{\bf H}/\partial t, \\ \nonumber
\nabla \times  {\bf H}&=&+\varepsilon_0\hat \varepsilon({\bf r},\omega)
\;\partial
 {\bf E}/\partial t,
\end{eqnarray}

\noindent
that obey Bloch's theorem. On-shell methods like the one
described in Ref.[8] are ideally suited to calculate the band
structure of materials whose dielectric function depends
on $\omega$, as graphite does.  
In these methods, $\omega$ is fixed at first and hence $\varepsilon(\omega)$ 
can be specified, 
allowing us to establish
an eigenvalue equation for the calculation of $k(\omega)$.
In particular this approach has a great advantage for studying   
EM waves in metals \cite{yo} where the dielectric function depends 
strongly on frequency.
Very recently \cite{Andrew}, an extension of this formalism
has been developed which is able to 
work with tensor like dielectric functions as the ones 
present in eqs. (4).  
Details of 
this formalism can be found in Refs. [8] and [16]. 

With this methodology we calculate all the allowed Bloch waves of our system
for a given photon
energy $\omega$. Then we can extract the dominant ones looking
at the imaginary parts of their wavevectors. Finally, we
can define the effective dielectric constants using  
Eq. (3). 

In its generalized version, MG approach is just the application 
of  Clausius-Mossotti relations to describe  
the dielectric properties of composite materials 
in an effective way. The basic 
assumption of this approach is that the polarizability of an individual 
object is modified by the presence of another ones only via 
their depolarizing contribution to the electric field acting 
on the object. For the case of arrays of cylinders in vacuum, 
a MG effective dielectric constant can be written as a 
function of the polarizability of an individual 
tube $\alpha_{\sigma}(\omega)$,  and  
a depolarization factor $L_{\sigma}$, 
which both depend on the polarization of the incident light,
$\sigma$: 

\begin{equation}
\varepsilon_{eff}^{\sigma , MG}(\omega)=1+ \frac{f \alpha_{\sigma}(\omega)}
{1-f L_{\sigma} \alpha_{\sigma}(\omega)},
\end{equation}
where $f$ is the volume fraction occupied by the cylinders. 

For p-polarized light $L_{p}=\frac{1}{2}$ and the  
polarizability of an individual plain cylinder is \cite{Lucas2} :

\begin{equation}
\alpha_p(\omega)=2 \frac{\varepsilon_{\parallel}(\omega)-\Delta}
{\varepsilon_{\parallel}(\omega)+\Delta},
\end{equation}
where $\Delta=\sqrt{\frac{\varepsilon_{\parallel}(\omega)}
{\varepsilon_{\perp}(\omega)}}$.  
Introducing this expression into  
Eq.(5), the effective dielectric constant for p-polarization 
can be written as:  


\begin{equation}
\varepsilon_{eff}^{p,MG}(\omega)=\frac{
\varepsilon_{\parallel}(\omega)+\Delta+f(
\varepsilon_{\parallel}(\omega)-\Delta)}
{\varepsilon_{\parallel}(\omega)+\Delta-f(
\varepsilon_{\parallel}(\omega)-\Delta)}.
\end{equation}

For s-polarized light, 
$L_s=0$ and the polarizability of a plain tube is just the same as 
the polarizability of the homogeneous medium, 
$\alpha_{s}(\omega)= 
\varepsilon_{\perp}(\omega)-1$, leading to a very simple 
formula for the effective dielectric constant:  

\begin{equation}
\varepsilon_{eff}^{s}(\omega)=f \varepsilon_{\perp}(\omega)+
(1-f).
\end{equation}

If we wish to consider the general case of hollow cylinders, 
$f$ must be replaced in 
eq. (8) by $f'=f(1-\gamma^2)$, $\gamma$ being 
the ratio between the inner and outer radii of the tubes.

Eq.(8) is the {\it exact} result for the 
effective dielectric constant for s-polarization. For this 
polarization there is no EM interaction between the tubes as 
MG approach assumes ($L_s=0$) and the total polarizability is 
just the sum of the polarizabilities of the individual 
nanotubes. Because the electric field is directed along the 
tubes, the optical response of the system depends only
on $\varepsilon_{\perp}(\omega)$. Our numerical results 
for this polarization are in complete agreement with Eq.(8), 
giving additional support to our scheme.

The simple relation (Eq.(8)) that holds for s-polarization  
can serve as an indirect way to estimate 
the volume fraction occupied by the nanotubes, $f$. 
The imaginary part of
$\varepsilon_{eff}^{s}$ is simply $f'$ multiplied  
by Im$\varepsilon_{\perp}$. This is the reason 
why the experimental Im$\varepsilon_{eff}^{s}$ 
\cite{waheer2} has its maximum 
at $4.6$ eV just as Im$\varepsilon_{\perp}$ does \cite{Palik}.
Then, a rough estimation of $f$ can be derived using the different 
heights of the peaks at $4.6$ eV for the reported Im$\varepsilon_{eff}^{s}$ 
and the experimental Im$\varepsilon_{\perp}$. Taking the inner radius 
of the tubes to be $0.25-2$ nm, we find $f \approx 0.6-0.7$.

For p-polarized light, 
the EM interaction between the nanotubes is not negligible and 
our numerical method can give new insights in the analysis of 
this interaction.
In figures 2 and 3 we show our numerical results for the effective dielectric
constant, comparing them with the results 
of the MG-like approach given by Eq.(7). 

With figure 2 our aim is to analyze the behaviour of these
dielectric constants for low volume fractions, $f$ varying 
from $f=0.09$ ($d=30$nm) to $f=0.27$ ($d=17$nm). As we can see 
from this figure,
the good agreement between MG results and 
our method is remarkable for these values of $f$; only at low frequencies there 
are some small differences in Im$\varepsilon_{eff}^{p}$ 
between both mean field theories. 
It is known that MG approach works very well 
in the dilute limit for dielectric media \cite{Datta}, 
and our numerical results show this 
is also the case for aligned carbon nanotubes. 
As we can conclude by looking at the two peaks 
present in Im$\varepsilon_{eff}^p (\omega)$, the 
optical response of the structure in the dilute limit 
is governed by the dipolar EM modes of an isolated 
carbon nanotube. Due to the anisotropy of the dielectric 
function of graphite, there are two dipolar modes, one  
located energetically at $5.3$ eV and the other at a higher 
energy, $6.3$ eV.

In figure 3 we study the case in which the EM interaction 
between the nanotubes becomes important,  
showing our results for 
the effective dielectric constant for distances between the tubes smaller
than three times their radii ($f > 0.3$). In the same figure 
we show the results of Eq. (7) for this range of distances and 
the experimental data for p-polarization as reported in Ref. \cite{waheer2}.
We can see that the effect of the
EM interaction is basically to shift to lower energies  
the low energy dipolar mode  
(from 5.3 eV for isolated nanotubes  
to 4.8 eV for very close contact nanotubes), and conceal the dipolar mode 
at $6.3$ eV. 
Again, MG approach seems to be a very good approximation 
for $\varepsilon_{eff}^{p}(\omega)$ for intermediate 
distances ($d > 11.5$nm) revealing that, for this range of distances,   
the EM coupling between the tubes is basically 
a dipolar one as MG assumes. However, the inclusion of the 
full EM interaction between the tubes as our mean field theory does 
is necessary in order to describe properly the 
strong coupling limit ($d \leq 11.5$nm). The experimental imaginary
part of the effective dielectric constant 
 presents its maximum just at $4.8$
eV, and our results suggest that the position of 
this peak can only be explained if the nanotubes are indeed in  
very close contact ($d \approx 10.3$nm) forming the films. 
As regards to the real part of $\varepsilon_{eff}^p$, the shape 
of the theoretical curve for $d=10.3$ nm closely matches 
the experimental data but a constant shift of approximately $1$
is needed in order to obtain an excellent   
agreement between theory and experiment.  

In conclusion, we have presented an effective medium 
approach to analyze the optical properties of carbon 
nanotube films. In contrast with usual effective 
medium methods, our numerical approach can treat the 
limit in which the 
EM interaction between the constituents of a complex 
system is important, as well as the dilute limit. 
We have also developed a Maxwell-Garnett approach 
to study this system. Using both mean field theories 
we have analyzed in detail the EM interaction between the 
nanotubes as a function of the volume fraction.
MG approach appears to be 
a good approximation for large and intermediate 
distances between the tubes. However, the inclusion 
of the full EM coupling between the nanotubes is essential in 
order to study closed-packed structures, appearing in 
aligned carbon nanotube films \cite{waheer2}.  
The good agreement obtained between 
our numerical results and the experimental data  
demonstrates that our mean field theory can be a very useful 
tool to analyze spectroscopic properties of 
nanostructured materials.

One of us (FJGV) acknowledges financial support of 
the Human Capital and Mobility Programme of the European Union under 
contract ERBFMBICT 950397 and useful discussions with Carmina 
Monreal and Fernando Flores. JMP gratefully acknowledges financial support of
the 
University of Basque Country, the Spanish Comisi\'on Asesora 
Cient\'{\i}fica y T\'ecnica (CICYT) and the Basque Unibertsitate eta Ikerketa Saila.
Part of this work was supported by EU
contract No. ERBCHRXCT 930342.

{\Large {\bf Figure Captions}} \newline

{\bf Figure 1}. a) Our model of a carbon nanotube film: an 
array of cylinders, infinitely long in the $z$-direction, is 
arranged on a square lattice in the xy-plane. 
b) Unit cell of the structure, 
$R=5$nm is the radius of the cylinders, $d$ is the distance between 
the centers of the tubes and $L_y$ is the thickness in the 
direction of propagation of the light that will be chosen equal to $d$. 
Also shown are the cylindrical unit vectors, $\hat{\theta}$ and 
$\hat{r}$ in order to describe the change of coordinates  
from cylindrical to Cartesians and the subsequent 
change in the tensor like dielectric function associated to  
a generic point inside the tube (x,y) (see text).  \newline   

{\bf Figure 2}. The real and imaginary parts of the 
effective dielectric constant for p-polarized light,  
with $d$ varying from $d=30$nm to 
$d=17$nm. Full line: our numerical results. Dashed line: results 
using a MG approach (see eq.7). 
\newline 

{\bf Figure 3}. The real and imaginary parts of the 
effective dielectric constant for p-polarized light  
for different 
values of $d$. 
(a) $d=10.3$nm (b) $d=10.7$nm (c) $d=11.5$nm 
(d) $d=12.5$nm (e) $d=14$nm.   
Full line: our numerical results. Dashed line: results 
using a MG approach (see eq.7). The dots represent the 
experimental data as reported in \cite{waheer2}. For the 
sake of clarity, the curves for the real part are shifted 
by a quantity shown in the figure. The experimental values 
of $Re \varepsilon_{eff}^{p}$ are shifted by a constant value of $+5$ with 
respect to the reported ones.

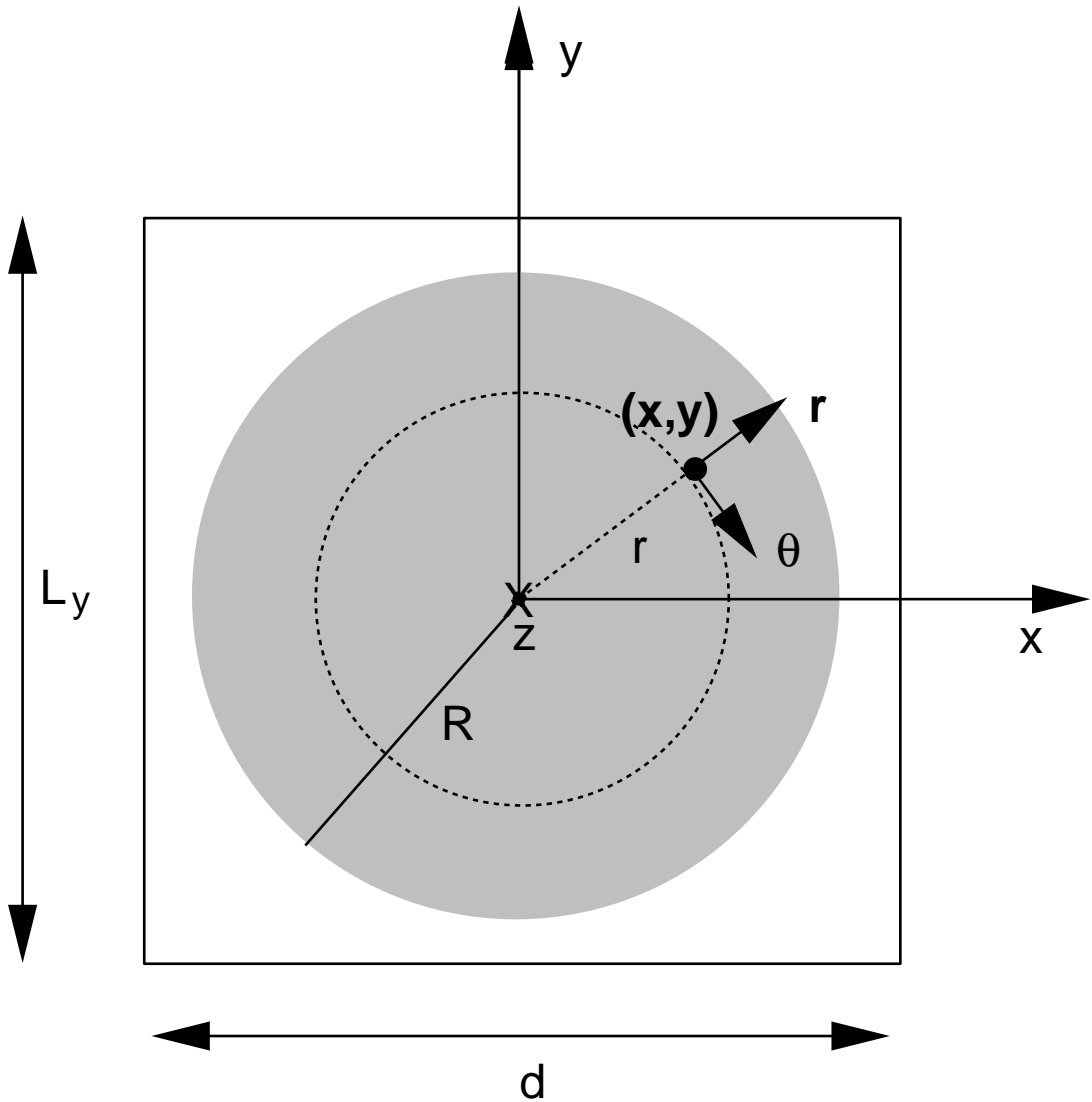
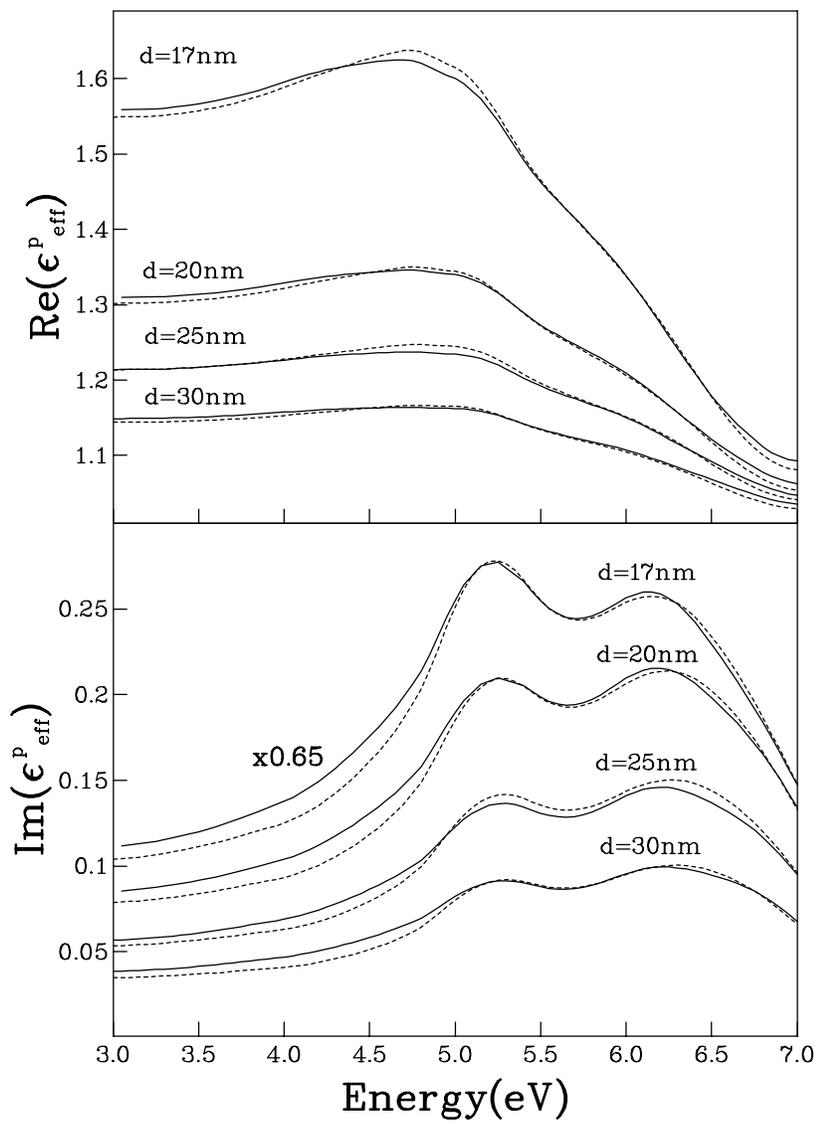
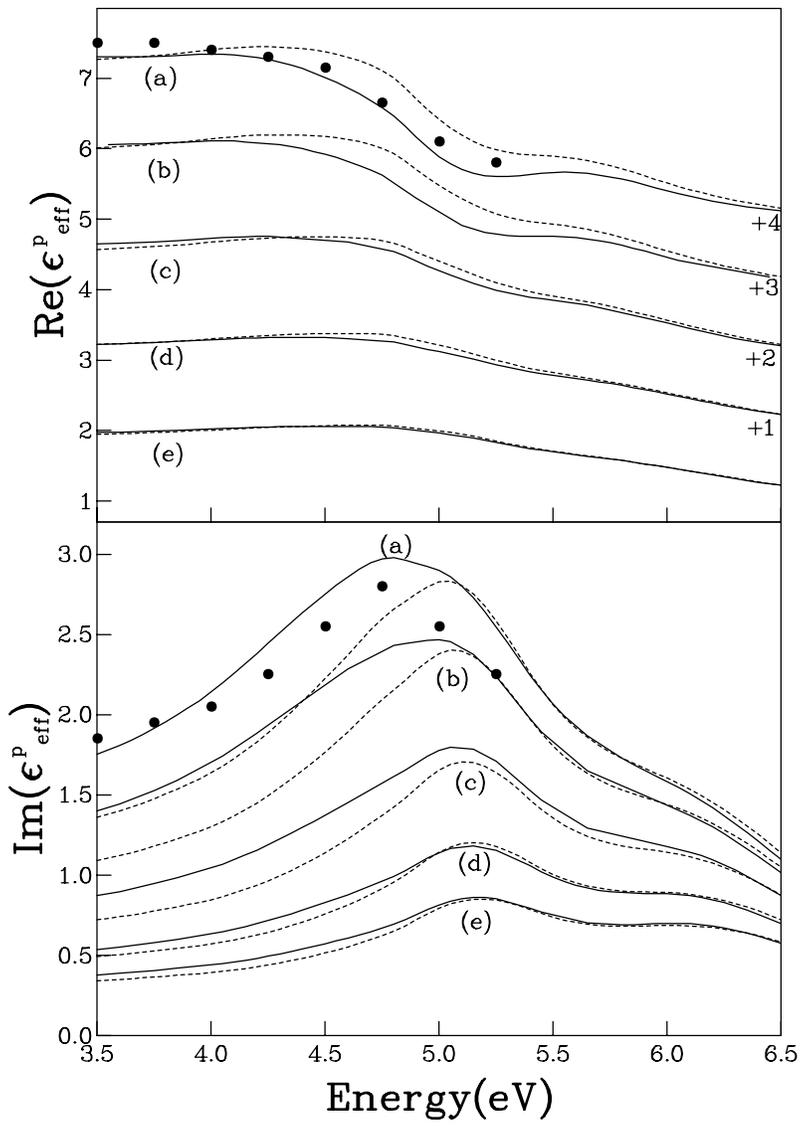
\end{document}